# Solution of the Dirac equation with position-dependent mass in the Coulomb field

A. D. Alhaidari

*Physics Department, King Fahd University of Petroleum & Minerals, Box 5047, Dhahran 31261, Saudi Arabia*
e-mail: haidari@mailaps.org

We obtain exact solution of the Dirac equation for a charged particle with position-dependent mass in the Coulomb field. The effective mass of the spinor has a relativistic component which is proportional to the square of the Compton wavelength and varies as $1/r$. It is suggested that this model could be used as a tool in the renormalization of ultraviolet divergences in field theory. The discrete energy spectrum and spinor wave-function are obtained explicitly.

## I. INTRODUCTION

Systems with spatially dependent effective mass were found to be very useful in studying the physical properties of various microstructures. Special applications in condensed matter are found in the investigation of electronic properties of semiconductors [1], quantum wells and quantum dots [2], $^3$He clusters [3], quantum liquids [4], graded alloys and semiconductor heterostructures [5], etc. These applications stimulated the development of various methods and techniques for studying systems with mass that depends on position. Recently, several contributions have emerged that give solutions of the wave equation for such systems. The one dimensional Schrödinger equation with smooth mass and potential steps was solved exactly by Dekar *et al.* [6]. The formalism of supersymmetric quantum mechanics was extended to include position dependent mass [7]. Shape invariance was also addressed in this setting and the energy spectra were obtained for several examples. A class of solutions was obtained explicitly for such systems with equi-spaced spectra [8]. Coordinate transformations in super-symmetric quantum mechanics were used to generate isospectral potentials for systems with position-dependent mass [9]. The ordering ambiguity of the mass and momentum operators and its effect on the exact solutions was addressed by de Souza and Almeida [10] where several examples are considered. so(2,1) Lie algebra as a spectrum generating algebra and as a potential algebra was used to obtain exact solutions of the effective mass wave equation [11]. Point canonical transformation (PCT) was used to obtain the energy spectra and wavefunctions for a large class of problems in one and three dimensions [12]. A class of quasi-exactly solvable problems with effective mass was presented by Koç *et al.* [13] where the wavefunctions are obtained in terms of orthogonal polynomials satisfying mass dependent recurrence relation. The algebraic method utilizing a given realization of the generators of the su(1,1) algebra was used in [14] to generate exactly solvable systems with position-dependent mass. The nonrelativistic Green's function was formally addressed in [15] by using path integral formulation to relate the constant mass Green's function to that of position-dependent mass. Quite recently, the PCT method developed in [12] was extended to give an explicit construction of the two-point Green's function for various potential classes of systems with position-dependent mass [16].

The relativistic extension of these formulations, on the other hand, remains undeveloped. One of the advantages of such an extension is the elimination of the ordering ambiguity of the mass and momentum operators which is present in the nonrelativistic kinetic energy $\frac{1}{2m}P^2$. This issue was addressed by Cavalcante *et al.* [17]. In this article we obtain a solution of the three dimensional Dirac equation with the Coulomb potential for a particle with position-dependent mass. The discrete energy spectrum and spinor wavefunction are obtained. In atomic units ($\hbar = m_0 = 1$), we take the following spherically symmetric singular mass distribution

$$m(r) = 1 + m\lambdabar^2/r \tag{1.1}$$

where $\lambdabar$ is the Compton wavelength $\hbar/m_0 c = c^{-1}$, and $m$ is a real scale parameter with inverse length dimension. The rest mass of the particle ($m_0 = 1$) is obtained either as the asymptotic limit ($r \to \infty$), or the nonrelativistic limit ($\lambdabar \to 0$) of $m(r)$. Consequently, a possible interpretation for this singular mass term might be found in relativistic quantum field theory. It could be considered as a model contributing to the "renormalization" of ultraviolet divergences which occur at high energies (equivalently, small distances $r \to 0$). In such a model, these divergences are renormalized into the singular mass term where $m$ stands for the renormalization scale. It should also be noted that this position-dependent mass term has a relativistic origin as well since it is proportional to the Compton wavelength which vanishes as $c \to \infty$ (equivalently, $\lambdabar \to 0$).

In the following section, we set up the three dimensional Dirac equation for a spinor with position-dependent mass interacting with the electromagnetic 4-potential $(A_0, \vec{A})$. We impose spherical symmetry and consider the special case where the space component of the electromagnetic potential vanishes (i.e., $\vec{A} = 0$). The time component, on the other hand, is taken as the Coulomb potential. That is $A_0 = \lambdabar V(r)$, where $V(r) = Z/r$ and $Z$ is the particle charge in units of $e$. The problem is then reduced to solving the radial component of the Dirac equation. A global unitary transformation is applied to this equation to separate the variables such that the resulting second order radial differential equation for the two spinor components become Schrödinger-like. This makes the solution of the relativistic problem easily attainable by simple and straightforward correspondence with the well-known exactly solvable nonrelativistic problem. The correspondence results in a map among the relativistic and nonrelativistic parameters. Using this map and the known nonrelativistic energy spectrum one can easily and directly obtain the relativistic spectrum. Moreover, the two radial components of the spinor wavefunction are obtained from the nonrelativistic wavefunction using the same parameter map.

## II. ENERGY SPECTRUM AND WAVEFUNCTION

Dirac equation is a relativistically invariant first order differential equation in 4-dimensional space-time for a four-component wavefunction ("spinor") $\psi$. For a free structureless particle it reads $(i\hbar \gamma^m \partial_m - Mc)\psi = 0$, where $M$ is the mass of the particle and $c$ the speed of light. The summation convention over repeated indices is used. That is, $\gamma^m \partial_m \equiv \sum_{m=0}^{3} \gamma^m \partial_m = \gamma^0 \partial_0 + \vec{\gamma} \cdot \vec{\partial} = \gamma^0 \frac{\partial}{c\partial t} + \vec{\gamma} \cdot \vec{\nabla}$. $\{\gamma^m\}_{m=0}^{3}$ are four constant square matrices satisfying the anticommutation relation $\{\gamma^m, \gamma^n\} = \gamma^m \gamma^n + \gamma^n \gamma^m = 2\eta^{mn}$, where $h$ is the metric of Minkowski space-time which is equal to $\text{diag}(+,-,-,-)$. A four-dimensional matrix representation that satisfies this relation is as follows:

$$g^0 = \begin{pmatrix} I & 0 \\ 0 & -I \end{pmatrix}, \quad \vec{g} = \begin{pmatrix} 0 & \vec{s} \\ -\vec{s} & 0 \end{pmatrix} \tag{2.1}$$

where $I$ is the 2×2 unit matrix and $\vec{s}$ are the three 2×2 hermitian Pauli matrices. The mass $M$ is generally space-time dependent in which case it should transform as a scalar function under the action of the Lorentz transformation. In atomic units ($\hbar = m_0 = e = 1$), $\lambdabar = 1/c$ and Dirac equation reads $\left(ig^m \partial_m - m/\lambdabar\right)\psi = 0$, where $m = M/m_0 = m(t, \vec{r})$. Next, we let the Dirac spinor be charged and coupled to the 4-component electromagnetic potential $A_m = (A_0, \vec{A})$. Gauge invariant coupling, which is accomplished by the "minimal" substitution $\partial_m \to \partial_m + iA_m$, transforms the free Dirac equation to $\left[ig^m(\partial_m + iA_m) - m/\lambdabar\right]\psi = 0$ which, when written in details, reads as follows

$$i\lambdabar \frac{\partial}{\partial t}\psi = \left(-i\vec{a}\cdot\vec{\nabla} + \vec{a}\cdot\vec{A} + A_0 + \frac{m}{\lambdabar}b\right)\psi \equiv \lambdabar^{-1}H\psi \tag{2.2}$$

where $H$ is the Hamiltonian, $\vec{a}$ and $b$ are the hermitian matrices defined as

$$\vec{a} = g^0\vec{g} = \begin{pmatrix} 0 & \vec{s} \\ \vec{s} & 0 \end{pmatrix} \text{ and } b = g^0 = \begin{pmatrix} I & 0 \\ 0 & -I \end{pmatrix} \tag{2.3}$$

Substituting these in Eq. (2.2) gives the following matrix representation of the Hamiltonian

$$H = \begin{pmatrix} m + \lambdabar A_0 & -\lambdabar i\vec{s}\cdot\vec{\nabla} + \lambdabar\vec{s}\cdot\vec{A} \\ -\lambdabar i\vec{s}\cdot\vec{\nabla} + \lambdabar\vec{s}\cdot\vec{A} & -m + \lambdabar A_0 \end{pmatrix} \tag{2.4}$$

Thus the eigenvalue wave equation reads $(H - \varepsilon)\psi = 0$, where $\varepsilon$ is the relativistic energy which is real. Now, we choose $\vec{A} = 0$ and impose spherical symmetry by taking $A_0 = \lambdabar V(r)$ and $m = m(r)$. In this case, the angular variables could be separated and we can write the spinor wavefunction as [18]

$$\psi = \begin{pmatrix} i[g(r)/r]\chi_{\ell m}^j \\ [f(r)/r]\vec{s}\cdot\hat{r}\chi_{\ell m}^j \end{pmatrix} \tag{2.5}$$

where $f$ and $g$ are real radial functions, $\hat{r}$ is the radial unit vector, and the angular wavefunction for the two-component spinor is written as

$$\chi_{\ell m}^j(\Omega) = \frac{1}{\sqrt{2\ell+1}} \begin{pmatrix} \sqrt{\ell \pm m + 1/2}\ Y_\ell^{m-1/2} \\ \pm\sqrt{\ell \mp m + 1/2}\ Y_\ell^{m+1/2} \end{pmatrix}, \quad \text{for } j = \ell \pm \tfrac{1}{2} \tag{2.6}$$

$Y_\ell^{m\pm 1/2}$ is the spherical harmonic function. In Eqs. (2.5) and (2.6), the letter $m$ stands for the integers in the range $-\ell, -\ell+1, ..., \ell$ and should not be confused with the position-dependent mass $m(r)$. Spherical symmetry gives $i\vec{s}\cdot(\vec{r}\times\vec{\nabla})\psi(r,\Omega) = -(1+\kappa)\psi(r,\Omega)$, where $\kappa$ is the spin-orbit quantum number defined as $\kappa = \pm(j+\tfrac{1}{2}) = \pm 1, \pm 2, ...$ for $\ell = j \pm \tfrac{1}{2}$. Using this we obtain the following useful relations

$$(\vec{s}\cdot\vec{\nabla})(\vec{s}\cdot\hat{r})F(r)\chi_{\ell m}^j = \left(\frac{dF}{dr} + \frac{1-\kappa}{r}F\right)\chi_{\ell m}^j$$

$$(\vec{s}\cdot\vec{\nabla})F(r)\chi_{\ell m}^j = \left(\frac{dF}{dr} + \frac{1+\kappa}{r}F\right)(\vec{s}\cdot\hat{r})\chi_{\ell m}^j \tag{2.7}$$

Employing these in the wave equation $(H - e)\mathbf{y} = 0$ results in the following 2×2 matrix equation for the two radial spinor components

$$\begin{pmatrix} m(r) + \lambdabar^2 V(r) - e & \lambdabar\left(\frac{\mathbf{k}}{r} - \frac{d}{dr}\right) \\ \lambdabar\left(\frac{\mathbf{k}}{r} + \frac{d}{dr}\right) & -m(r) + \lambdabar^2 V(r) - e \end{pmatrix} \begin{pmatrix} g(r) \\ f(r) \end{pmatrix} = 0 \qquad (2.8)$$

Now, we specialize further to the Coulomb potential $V(r) = Z/r$ and take $m(r)$ as the position dependent mass given by Eq. (1.1). This maps Eq. (2.8) into the following

$$\begin{pmatrix} 1 + \lambdabar^2 \frac{Z+\mathbf{m}}{r} - e & \lambdabar\left(\frac{\mathbf{k}}{r} - \frac{d}{dr}\right) \\ \lambdabar\left(\frac{\mathbf{k}}{r} + \frac{d}{dr}\right) & -1 + \lambdabar^2 \frac{Z-\mathbf{m}}{r} - e \end{pmatrix} \begin{pmatrix} g \\ f \end{pmatrix} = 0 \qquad (2.9)$$

which results in two coupled first order differential equations for the two radial spinor components $f$ and $g$. Eliminating one component in favor of the other gives a second order differential equation. This equation is not Schrödinger-like (i.e., it contains first order derivatives). To obtain a Schrödinger-like equation we proceed as follows. A global unitary transformation $U(\mathbf{h}) = \exp(\frac{i}{2}\lambdabar \mathbf{h s}_2)$ is applied to the radial Dirac equation (2.9), where $\mathbf{h}$ is a real constant parameter and $\mathbf{s}_2$ is the 2×2 Pauli matrix $\begin{pmatrix} 0 & -i \\ i & 0 \end{pmatrix}$. The Schrödinger-like requirement dictates that the parameter $\mathbf{h}$ satisfies the constraint:

$$C\mathbf{m} + S\mathbf{k}/\lambdabar = \pm Z \qquad (2.10)$$

where $S = \sin(\lambdabar \mathbf{h})$, $C = \cos(\lambdabar \mathbf{h})$ and $-\frac{\pi}{2} \le \lambdabar \mathbf{h} \le +\frac{\pi}{2}$. The solution of this constraint gives two angles whose cosines are

$$C = \left(\mathbf{m}^2 + \mathbf{k}^2/\lambdabar^2\right)^{-1} \left[\pm \mathbf{m} Z + t \frac{|\mathbf{k}|}{\lambdabar} \sqrt{\left(\frac{\mathbf{k}}{\lambdabar}\right)^2 + \mathbf{m}^2 - Z^2}\right] > 0 \qquad (2.11)$$

where $t$ is a $\pm$ sign which is chosen such that the resulting value is positive. We choose not to fix the sign in Eq. (2.10) but continue the development with the two possibilities. Consequently, one should distinguish between two values for the angular parameters $C$ and $S$ (e.g., $C_\pm$ and $S_\pm$). However, for economy and simplicity of notation we will maintain the same symbols. Equation (2.9) is now transformed into the following

$$\begin{pmatrix} C - e + (1\pm 1)\lambdabar^2 \frac{Z}{r} & \lambdabar\left(-\frac{S}{\lambdabar} + \frac{\mathbf{g}}{r} - \frac{d}{dr}\right) \\ \lambdabar\left(-\frac{S}{\lambdabar} + \frac{\mathbf{g}}{r} + \frac{d}{dr}\right) & -C - e + (1\mp 1)\lambdabar^2 \frac{Z}{r} \end{pmatrix} \begin{pmatrix} f^+(r) \\ f^-(r) \end{pmatrix} = 0 \qquad (2.12)$$

where $\mathbf{g} = t\frac{|\mathbf{k}|}{\mathbf{k}}\sqrt{\mathbf{k}^2 + \lambdabar^2(\mathbf{m}^2 - Z^2)}$ and

$$\begin{pmatrix} f^+ \\ f^- \end{pmatrix} = U\mathbf{y} = \begin{pmatrix} \cos\frac{\lambdabar \mathbf{h}}{2} & \sin\frac{\lambdabar \mathbf{h}}{2} \\ -\sin\frac{\lambdabar \mathbf{h}}{2} & \cos\frac{\lambdabar \mathbf{h}}{2} \end{pmatrix} \begin{pmatrix} g \\ f \end{pmatrix} \qquad (2.13)$$

Equation (2.12) gives one spinor component in terms of the other as follows

$$f^\pm = \frac{\lambdabar}{C \mp e}\left(\pm\frac{S}{\lambdabar} \mp \frac{\mathbf{g}}{r} + \frac{d}{dr}\right) f^\mp \qquad (2.14)$$

Whereas, the resulting Schrödinger-like wave equation becomes

$$\left[-\frac{d^2}{dr^2} + \frac{\mathbf{g}(\mathbf{g}\pm 1)}{r^2} + 2\frac{Ze + \mathbf{m}}{r} - \frac{e^2 - 1}{\lambdabar^2}\right] f^\pm(r) = 0 \qquad (2.15)$$

Comparing this equation with that of the well-known nonrelativistic Coulomb problem with constant mass

$$\left[-\frac{d^2}{dr^2} + \frac{\ell(\ell+1)}{r^2} + 2\frac{Z}{r} - 2E\right]\Phi(r) = 0 \qquad (2.16)$$

gives, by correspondence, the following two maps between the parameters of the two problems:

for $f^+$: $\ell \to g$ or $\ell \to -g-1, Z \to Ze + m, E \to (e^2-1)/2\lambdabar^2$ (2.17a)

for $f^-$: $\ell \to g-1$ or $\ell \to -g, Z \to Ze + m, E \to (e^2-1)/2\lambdabar^2$ (2.17b)

The first (second) choice for $\ell$ in each one of these two maps is for positive (negative) values of $k$, respectively. Using these parameter maps in the well-known nonrelativistic energy spectrum, $E_n = -Z^2/2(\ell+n+1)^2$, gives the following relativistic spectrum

$$e_n^\ell = \left[1+\left(\frac{\lambdabar Z}{n+\ell+1}\right)^2\right]^{-1}\left[-\frac{\lambdabar^2 mZ}{(n+\ell+1)^2} \pm \sqrt{1+\lambdabar^2 \frac{Z^2-m^2}{(n+\ell+1)^2}}\right] \qquad (2.18)$$

where $\ell$ stands for either one of the four possible alternative values in (2.17) associated independently with $f^\pm$ and $\pm k \geq 1$. One can easily verify that in the "non-renormalization limit" ($m \to 0$) the familiar relativistic spectrum for the Coulomb problem with constant mass is recovered. Moreover, in the nonrelativistic limit ($\lambdabar \to 0$), the nonrelativistic spectrum is obtained. It is also interesting to note that in the absence of the Coulomb interaction ($Z = 0$) the position-dependent mass term is equivalent to a scalar potential $V_s(r) = m/r$ whose energy spectrum is obtained as $e_n^\ell = \pm\sqrt{1-\lambdabar^2 m^2/(n+\ell+1)^2}$. Now, the radial components of the spinor wavefunction is obtained using the same parameter map (2.17) and the nonrelativistic wavefunction

$$\Phi_n(r) = \sqrt{r_n \Gamma(n+1)/\Gamma(n+2\ell+2)}(r_n r)^{\ell+1} e^{-r_n r/2} L_n^{2\ell+1}(r_n r) \qquad (2.19)$$

where $r_n = 2|Z|/(n+\ell+1)$. The result of this map, under the compatibility relation (2.14), is the following

$$f_n^+(r) = \begin{cases} \sqrt{\frac{w_n^g \Gamma(n+1)}{\Gamma(n+2g+2)}}\sqrt{\frac{C+e_n^g}{2C}}(w_n^g r)^{g+1} e^{-w_n^g r/2} L_n^{2g+1}(w_n^g r) &, k>0 \\ \sqrt{\frac{w_n^{-g-1}\Gamma(n+1)}{\Gamma(n-2g)}}\sqrt{\frac{C+e_n^{-g-1}}{2C}}(w_n^{-g-1}r)^{-g} e^{-w_n^{-g-1}r/2} L_n^{-2g-1}(w_n^{-g-1}r) &, k<0 \end{cases} \qquad (2.20)$$

$$f_n^-(r) = \begin{cases} \sqrt{\frac{w_n^{g-1}\Gamma(n+1)}{\Gamma(n+2g)}}\sqrt{\frac{C-e_n^{g-1}}{2C}}(w_n^{g-1}r)^g e^{-w_n^{g-1}r/2} L_n^{2g-1}(w_n^{g-1}r) &, k>0 \\ \sqrt{\frac{w_n^{-g}\Gamma(n+1)}{\Gamma(n-2g+2)}}\sqrt{\frac{C-e_n^{-g}}{2C}}(w_n^{-g}r)^{-g+1} e^{-w_n^{-g}r/2} L_n^{-2g+1}(w_n^{-g}r) &, k<0 \end{cases} \qquad (2.21)$$

where $w_n^\ell = 2|Ze_n^\ell + m|/(n+\ell+1)$ and $C$ is given by Eq. (2.11) in which the $\pm$ sign goes with $f_n^\pm$, respectively. The two lowest energy states, where $e = \pm C$, are $y = \begin{pmatrix} 0 \\ f_0^- \end{pmatrix}$ for $k>0$ and $y = \begin{pmatrix} f_0^+ \\ 0 \end{pmatrix}$ for $k<0$.